\documentclass[9pt,twocolumn,twoside]{osajnl}

\journal{josab} 

\setboolean{shortarticle}{false} 

\usepackage{color}
\usepackage{upgreek}
\DeclareCaptionFormat{myformat}{#1#2#3\hrulefill}
\title{Role of short-range order in manipulating light absorption in disordered media}

\author[1]{M. Q. Liu}
\author[1,*]{C. Y. Zhao}
\author[1]{B. X. Wang}
\author[1]{Xing Fang}

\affil[1]{Institute of Engineering Thermophysics, School of Mechanical Engineering, Shanghai Jiao Tong University, Shanghai 200240, China}

\affil[*]{Corresponding author: Changying.zhao@sjtu.edu.cn}

\ociscodes{(290.0290) Multiple scattering; (160.0160) Metamaterials;(350.0350) Nanophotonics and photonic crystals, Solar energy.}

\begin{abstract}
Structural correlations have a significant effect on light propagation in disordered media. We numerically investigate the role of short-range order in light absorption in thin films with disordered nanoholes. Two types of disordered distributions, including stealthy hyperuniform (SHU) and hard disk (HD) patterns with different degrees of short-range order, are studied. We find that Bragg scattering induced by short-range order results in the appearance of a gradually sharper absorption peak with the increasing of degrees of short-range order ($\chi$, $\phi$). A physical model is proposed to calculate the in-plane angularly differential scattering cross section $d \sigma^*/d \theta$ of thin-film nanostructures with consideration of {the} structure factor $S(q)$. Results reveal that higher level of short-range order can enhance in-plane Bragg scattering in certain wavelengths and directions corresponding to rich and sharp peaks in {the} structure factor $S(q)$, which can further modify morphology-dependent-like resonances of an individual scatterer {and leads } to {large} improvement of absorptivity in thin films. Besides, the comparison results show that SHU structures exhibit better integrated absorption ($IA$) enhancement than both HD and periodic structures. And there is a transition of local-order phase between hexagonal lattice{s} and square lattice{s for SHU structures}, leading to an optimal absorption performance when $\chi$ is around 0.5 of interest. The present study paves a way in controlling light absorption and scattering using novel disordered nanostructures. 

\end{abstract}

\setboolean{displaycopyright}{true}

\begin{document}

\maketitle

\section{Introduction}
Controlling light transport and scattering in disordered photonic structures is crucial and challenging due to extremely complex interference phenomena of electromagnetic waves, which spawns a wide range of applications including advanced imaging techniques \cite{tanzid2016absorption}, quantum optics \cite{PhysRevLett.118.087203}, random lasing \cite{Jeong2012Lasing,Okamoto:17}, renewable energy \cite{aeschlimann2015perfect,Naqavi:13}, biophotonics and biomedical applications \cite{vcivzmar2011shaping}, etc. Compared to ordered ones, disordered media are not sensitive to fabrication imperfections, making them easier to fabricate in practice. Also, light can be scattered multiple times in complex structures and light-matter interactions can be efficiently enhanced. 

Moreover, disordered structures with controlled short-range order also provide possibilities to design versatile structures with specific photonic properties. {Recently, such structures have been used to study the generation of complete and isotropic photonic band gaps (PBGs) \cite{yang2010photonic,man2013photonic} in the absence of conventional Bragg scattering \cite{joannopoulos2011photonic} which has been seen as the precondition for PBGs in periodic structures.} Conley $et.~al$ \cite{conley2014light} focused on the characteristics of light transport and localization {phenomena} in 2D photonic structures with constraint structural correlations and the analysis revealed that tuning local order provides a feasible way to manipulate the transport mean free path and {the} Anderson localization length. Lasing action in biomimetic structures with short-range order also has been investigated experimentally and numerically. Results show that lasing performance can be improved and the lasing frequency can be easier manipulated via {the} structure factor \cite{PhysRevLett.106.183901}. Non-iridescent colors can also be obtained by controlling distributions of nanoparticles with different sizes and materials \cite{Dong:10,Xiaoe1701151}. Besides, some studies emphasized macroscopic physical phenomena \cite{aeschlimann2015perfect,riboli2011anderson}, including absorptivity of nanostructures. Amorphous structures exhibit widerband absorption performance \cite{yin2015high,fazio2016strongly,liu2017ultra} and absorptivity can be independent of incident {angles} \cite{fang2015thin,burresi2013two}, which are attractive especially for solar energy harvesting \cite{liu2017ultra,Zhong:15}.

Recently, the appearance of a new specific correlated structure called {stealthy hyperuniform} materials \cite{batten2008classical,torquato2003local}, in which the long-range density fluctuations {are} suppressed and the structural correlations can be well controlled by {the} constraint parameter $\chi$. These systems are developed in reciprocal space with {the} structure factor $S(k)=0$ vanishing with wavenumber $k<k_{\textup{c}}$. In other words, the number variance $\sigma^{2}_{\textup{N}}(R_{\textup{s}})\equiv \langle N(R_s)^{2} \rangle - \langle N(R_{\textup{s}}) \rangle ^{2}$ within a window of radius $R_{\textup{s}}$ grows much more slowly than the volume of the window $R^{d}_{\textup{s}}$ in real space, where $d$ denotes dimensionality. Larger value of $\chi$ indicates higher level of short-range order, where structures are more like ordered ones within short distance. {Due to the controllability of structural correlations and two or three dimensional media can be fabricated in practice \cite{man2013isotropic}, structures with SHU distributions have attracted much attention.} Interestingly, complete and isotropic PBGs can be developed in such media when the characteristic size {$a$, the averaged minimum distance between adjacent scatterers,} is comparable to {the} wavelength \cite{man2013isotropic,man2013photonic}, while it is difficult for periodic photonic crystals to induce PBGs for both polarizations simultaneously \cite{joannopoulos2011photonic}. {Besides}, disordered SHU structures can also be fully transparent in long waveband because of $S(k<k_{\textup{c}})=0$ where multiple scattering is chiefly suppressed \cite{leseur2016high}. A recent work \cite{froufe2016role} making comparison between {novel SHU and conventional HD structures} have focused on the formation of PBGs and the normalized density of states (NDOS) with different degrees of short-range order. Tailoring Bragg scattering in isotropic Brillouin zone by tuning the degree of local order, the width of PBGs and the distribution{s} of NDOS both in SHU and HD structures can be well controlled. Notably, high similarities in PBGs and NDOS indicate SHU structures can be seen as the counterparts of HD structures in reciprocal space \cite{froufe2016role}. Nevertheless, whether other macroscopic photonic characteristics of these two disordered structures, such as absorptivity, which is crucial in light trapping and energy harvesting, are similar or not is worth deeply exploring and further being discussed.

On the other hand, using periodic or random patterns to enhance absorptivity in thin films with two dimensional nanostructures has been demonstrated for many years \cite{Riboli2014Engineering,vynck2012photon}. Especially for solar energy harvesting, the thin-film nanostructures embedded with 2D patterns can support strong optical resonances, leading to efficient light trapping and radiation absorption enhancement \cite{Callahan2012Solar}. Most of the theoretical and experimental investigations in thin-film nanostructures have been focused on periodic structures, for instance, nanoholes and nanowire arrays with triangular and square lattices \cite{Bao:10,Basu2012Ultrathin}. {Also}, theoretical analysis proposed by Yu $et. ~al$ \cite{Yu2010Fundamental} illustrated that random conditions can be helpful to achieve broadband absorption enhancement{, which has been numerically and experimentally verified \cite{fang2015thin,2013Light,bao2012investigation,Vynck2012Disordered}}. {However, }the random conditions studied before \cite{Vynck2012Disordered,vynck2012photon} are arbitrary to some extent, not only making the absorptivity manipulation to be much more difficult but also lacking systematic exploration about underlying radiation absorption mechanisms. {Besides}, amorphous photonic structures, which possess short-range order, are widely discovered in the nature, the eyes and feather of birds for example, showing a promising potential in bionics applications in practice \cite{Siddiquee1700232,Xiaoe1701151}. Therefore, it is necessary to study the influence of structures with controlled short-range order in absorptivity and the underlying physical mechanisms should {also} be investigated in detail.      

Here in this paper we will demonstrate that SHU structures with different levels of short-range order are capable to achieve better absorption enhancement by comparing to traditional HD structures. To explore the absorptivity of thin-film structures with short-range order, we have carried out {numerical} simulations on supercells with periodic boundary conditions by {using} finite difference time-domain method. Analysis {about the size effects} ensures us to {restore the short-range order in a supercell.} Then we focus on the influence of different levels of short-range order in absorptivity. {By studying the in-plane effective differential scattering cross sections considering the influence of positional correlations, we show that Bragg scattering induced by local order, plays a substantial role in strengthening in-plane scattering and enhancing absorption.}

{\tiny }

\section{Model and methods}
\begin{figure}[tbp]
	\centering
	\includegraphics[width=1.1\linewidth]{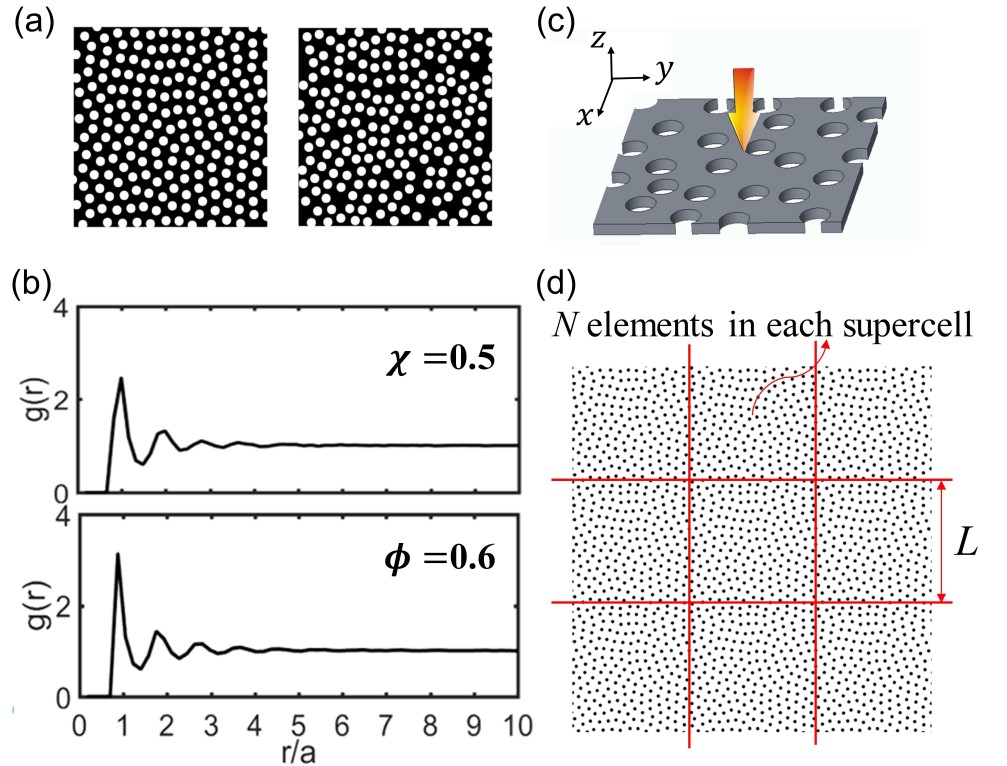}
	\caption{$\mathbf{(a)}$ Typical configurations of disordered structures for SHU patterns (left, $\chi$=0.5) and HD patterns (right, $\phi$=0.6) with $\mathbf{(b)}$ corresponding pair distribution functions  $g(r)$ (top: SHU patterns; bottom: HD patterns). The positions of several pronounced peaks appearing in small $r$ are similar for both structures. $\mathbf{(c)}$ Schematic of the thin film with disordered nanoholes. $\mathbf{(d)}$ Periodic boundary conditions are considered when performing numerical calculations.}\label{fig:1} 
\end{figure}

In this work, two algorithms \cite{froufe2016role} are adopted to obtain positions of disordered patterns in two-dimensional systems, including {SHU and HD }patterns with different degrees of structural correlations as measured by $\chi$ and $\phi$, respectively. Disordered HD patterns are generated by the isochoric Monte Carlo simulation with periodic boundary conditions. SHU patterns are obtained by combining molecular dynamics simulation and a simulated annealing relaxation scheme running a system of particles with an optimal pair energy potential \cite{PhysRevX.5.021020}
\begin{equation}
E=\sum_{|\mathbf{k}|\le k_{\textup{c}}}S(\mathbf{k})=\frac{1}{N}\sum_{|\mathbf{k}|\le k_{\textup{c}}}\big\|\sum_{j=1}^N e^{{-i\mathbf{k}}\cdot{\mathbf{R}_{j}}}\big\|^{2},
\end{equation}
in which $\mathbf{k}$ is {the} wavevector, $N$ is the number of particles and $\mathbf{R}_{j}$ is the position of the $j$-th particle. The level of short-range order of SHU patterns is usually measured by $\chi=\frac{M(k_{\textup{c}})}{2(N-1)}$ in a 2D system \cite{froufe2016role}, where $M(k_{\textup{c}})$ counts the number of independently constrained wavevectors \cite{PhysRevX.5.021020}. Note that  for simplicity we fix $k_c=\sqrt{(2M(k_{\textup{c}})+1)/\pi}$ in this paper, the side length of the constrained region $\Omega$, which is instructive in predicting size effects of SHU structures. Therefore, disordered distributions are controlled with the target number $N$ of seed points and the constraint parameters $\chi$ for SHU patterns and $\phi$ for HD patterns. Increasing $\phi$ and $\chi$ results in significant peaks in pair distribution function $g(r)$, indicating the development of local positional order, which will be discussed in Sec. {3}. The $g(r)$ measures the statistical averaged probability of finding another scatterer at a given distance $r$ from the center scatterer normalized by the average number density $\rho=N/L^2$ ($L$ is the size of the supercell) and gives 

\begin{equation}
g(r)=\frac{1}{\rho}\frac{P(r)}{s_1(r)dr},
\end{equation}
in which $P(r)$ is the probability and $s_1(r)$ is the surface area of a single sphere of radius $r$. Figure \ref{fig:1}(a) shows typical SHU and HD patterns at $\chi=0.5$ (left) and $\phi=0.6$ (right) with corresponding pair distribution functions $g(r)$ (Fig. \ref{fig:1}(b)), respectively. Note that $g(r)=1$ reveals that the point distribution in {one} system is totally random. Apparently, long-range density fluctuation{s are} both suppressed in two types of patterns, and each pattern possesses a similar short-range order shown in Fig. \ref{fig:1}(b). Interestingly, here in this paper we will show the optical absorptivity of two structures are {different} in spite of similarity. {Note that the $x$-axis of $g(r)$ is normalized by characteristic length $a=1/\sqrt{\rho}$ showing distinct statistical features of structural correlations, which will be discussed in detail in the next part.}

\begin{figure}[t]
	\centering
	\includegraphics[width=0.9\linewidth]{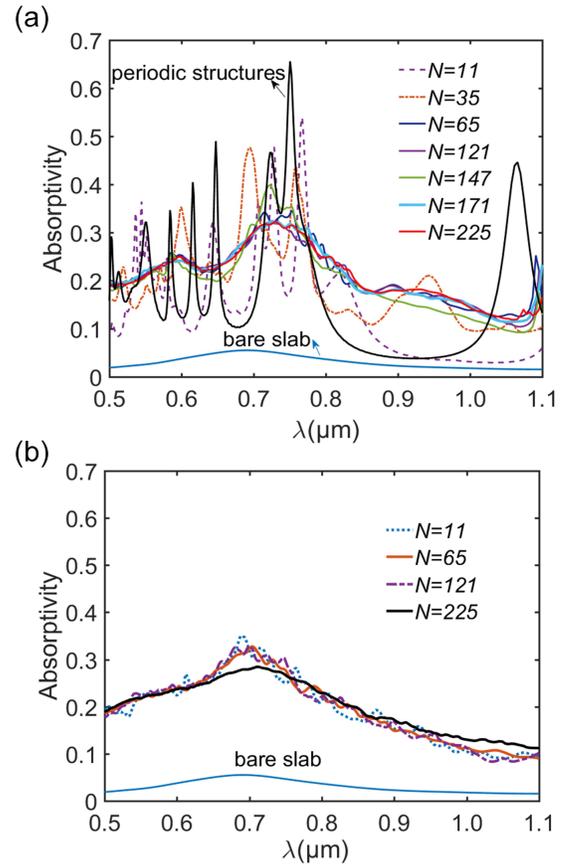}
	\caption{{The absorption spectra of~$\mathbf{(a)}$ SHU structures and~$\mathbf{(b)}$ HD structures {change} with the number of nanoholes.}}\label{fig:2}
\end{figure}

After all cases of disordered patterns are prepared, a thin film embedded with circular air holes with disordered distributions {(see Fig.~\ref{fig:1}(c))} is considered to investigate the absorptivity of a nanostructured film compared with that of the bare film. The three-dimensional finite difference time-domain (FDTD solutions, Lumerical) is applied to perform numerical calculations. In order to explore the capability of disordered structures in enhancing absorption, we study a fictitious material (similar to {c}-Si) with very weak intrinsic absorption, whose real part of permittivity is 12 and the absorption length is 3.3 $\upmu$m {over the entire incident spectrum}, {which are adopted from Ref. \cite{vynck2012photon,fang2015thin}.} The thickness of the thin film is designed to be 0.1 $\upmu$m, {which is the typical size in thin-film solar cells {\cite{Pratesi2013Disordered}}. The constant radius of air holes and the filling fraction are chosen to be $R= 0.18$ $\upmu$m and $f=N\pi R^2/L^2=0.3$, {which correspond to optimal parameters of square-lattice ordered structures with better absorption performance \cite{2013Light} }. In the simulation region, periodic boundary conditions (see Fig.~\ref{fig:1}(d)) are applied in the $x$ and $y$ directions, while we employ perfectly matched layer (PML) condition in the normal direction. A plane wave with $y$-polarization under normal incidence from the vacuum side is considered ranging from 0.5 $\upmu$m to 1.1 $\upmu$m, and two normalized power monitors are placed above and below the slab along the $z$ direction to extract the information about reflectance $R$ and transmittance $T$. Therefore, the fundamental absorptivity can be analyzed by energy conservation $A=1-R-T$.

\section{Effect of supercell size} 

As usual, many studies have focused on high-density hyperuniform materials with large sizes \cite{leseur2016high,zhang2016transport}, whose statistical short-range ordered features of point distributions can be easily achieved. {It} seems to be time-consuming and not applicable to investigate the absorption features with so large configurations where the number of  scatterers is larger than 1000. However, for smaller configurations, the size effects of the supercell we consider here (see the red square in Fig. \ref{fig:1}(d)) cannot be ignored since periodic boundary conditions are applied in $x$ and $y$ directions {\cite{Martins2012Engineering}}. Therefore, it is necessary to find {a proper} size of the supercell which can fully achieve the main feature of SHU structures with desirable short-range order before investigating the influence of different levels of structural correlations in optical absorptivity. The nanostructured thin films containing different numbers of air holes ranging from 11 to 225 are considered in this part. For different cases, the area of the slab increases simultaneously to ensure that the filling fraction ($f=0.3$) and constraint parameters ($\phi=0.6$, $\chi=0.5$) are unchanged. Notably, we have confirmed that HD structures with $\phi=0.6$ and SHU structures with $\chi$=0.5 show high similarity in $g(r)$ (Figs. \ref{fig:1}(a), \ref{fig:1}(b)). The absorption spectra are obtained by calculating {50} configurations to get ensemble average data.
\begin{figure}[tb]
	\centering
	\includegraphics[width=0.89\linewidth]{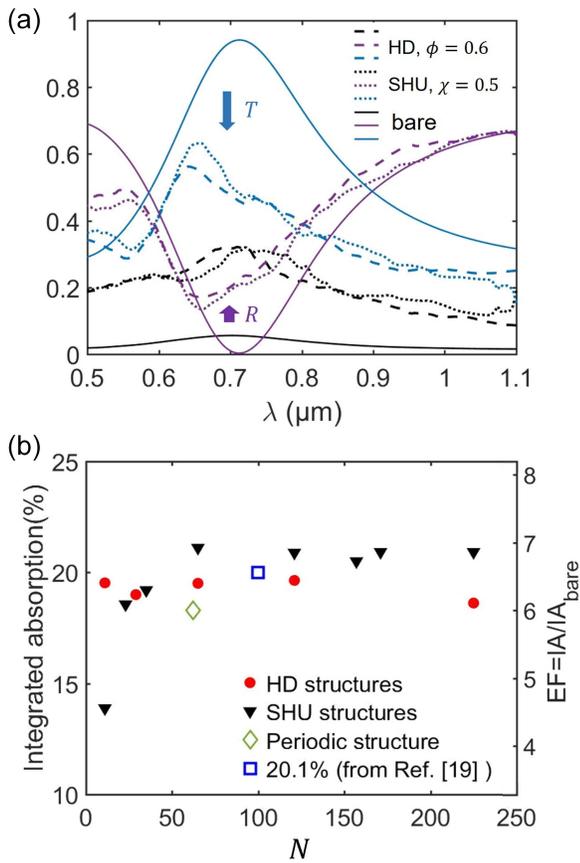}
	\caption{{$\mathbf{(a)}$ The radiation spectra including $A$ (black lines), $R$ (purple lines) and $T$ (blue lines) for HD structures ($\phi=0.6$, $N = 225$) and SHU structures ($\chi=0.5$, $N = 225$) compared with the bare slab.} {$\mathbf{(b)}$ The integrated absorption as a function of the number of nanoholes for two disordered structures.} $EF={IA}/{IA_{\textup{bare}}}$. The $EF$ of the {periodic structure} is smaller than that of SHU and HD structures when the values of integrated absorption keep steady, suggesting the amorphous structures are superior to ordered counterparts.} \label{fig:3}
\end{figure}

The results are illustrated in {Fig.~\ref{fig:2}} with some remarkable differences between two disordered structures. Firstly, the thin films embedded with disordered nanoholes show distinct absorption enhancement compared to that of the bare slab exhibiting a fairly broad, low-$Q$ ($Q$ is the quality factor) peak centered at 0.7 $\upmu$m induced by the Fabry-Perot resonance \cite{vynck2012photon}. { The reflection and transmission spectra when $N$ = 225 are illustrated in Fig. \ref{fig:3}(a) as well. Compared with the results of the bare slab, the transmittance for both disordered structures drops sharply by about 50$\%$ and the reflectance is enhanced and broadened at the same time, leading to broadband absorption enhancement. \cite{Muller2014Photonic}} {Besides, as shown in Fig. \ref{fig:2}, }HD structures are insensitive to the {sizes} and a broad absorption peak is well-preserved for every case, while structures with SHU patterns present rich spectral features with several resonances over the entire working wavelength especially for cases with small domain area. In particular, for cases with 11 nanoholes, there are several sharp interference peaks appearing around short wavelength, implying the development of a pseudo-order structure. Then the absorption spectra become much smoother with the increasing of the holes' number albeit the integrated absorption $(IA)$ does not decrease.

Consistently, the integrated absorption of two structures with the same constrained local order are diverse as well. Here, we calculate the integrated absorption defined as 
\begin{equation}
{IA}=\frac{1}{\lambda_{\textup{max}}-\lambda_{\textup{min}}} \int_{\lambda_{\textup{min}}}^{\lambda_{\textup{max}}} A(\lambda) d \lambda,
\end{equation} \label{Eq:2}
where $\lambda$ is {the} wavelength and subscripts $max$ and $min$ represent the maximum and minimum wavelength of interest, respectively. To make a clear comparison with the bare slab, we define the enhancement factor $EF={IA}/{IA_{\textup{bare}}}$, which is the ratio of the integrated absorption of disordered structures to that of the bare slab.~As shown in {Fig.~\ref{fig:3}(b)}, both HD and SHU structures, superior to the ordered one, can enhance absorption greatly with $EF>6$. {The maximum $IA$ of SHU structures when absorptivity keeps steady is larger than results of amorphous structures studied in Ref. \cite{fang2015thin}, where the working wavelength of interest and the optical parameters of materials are similar to ours. Because the disordered conditions in that paper are randomly chosen, our well-designed disordered structures show better absorption performance.} {Moreover,} the $IA$ of HD structures does not vary substantially and keep steady at about 19\% regardless of the variation of the number of holes, which is consistent with the absorption spectra in Fig.~\ref{fig:2}(b). {In contrast}, for SHU structures, the amplitude of $EF$ increases, and saturates in {a} limit of the large systems after the number of holes reaching over 60 with slight fluctuations. It indicates that there is a minimum size to minimize the size effects of SHU structures, and the absorption performance is predicted to be steady when approaching this value.

\begin{figure}[tbp]
	\centering
	\includegraphics[width=1\linewidth]{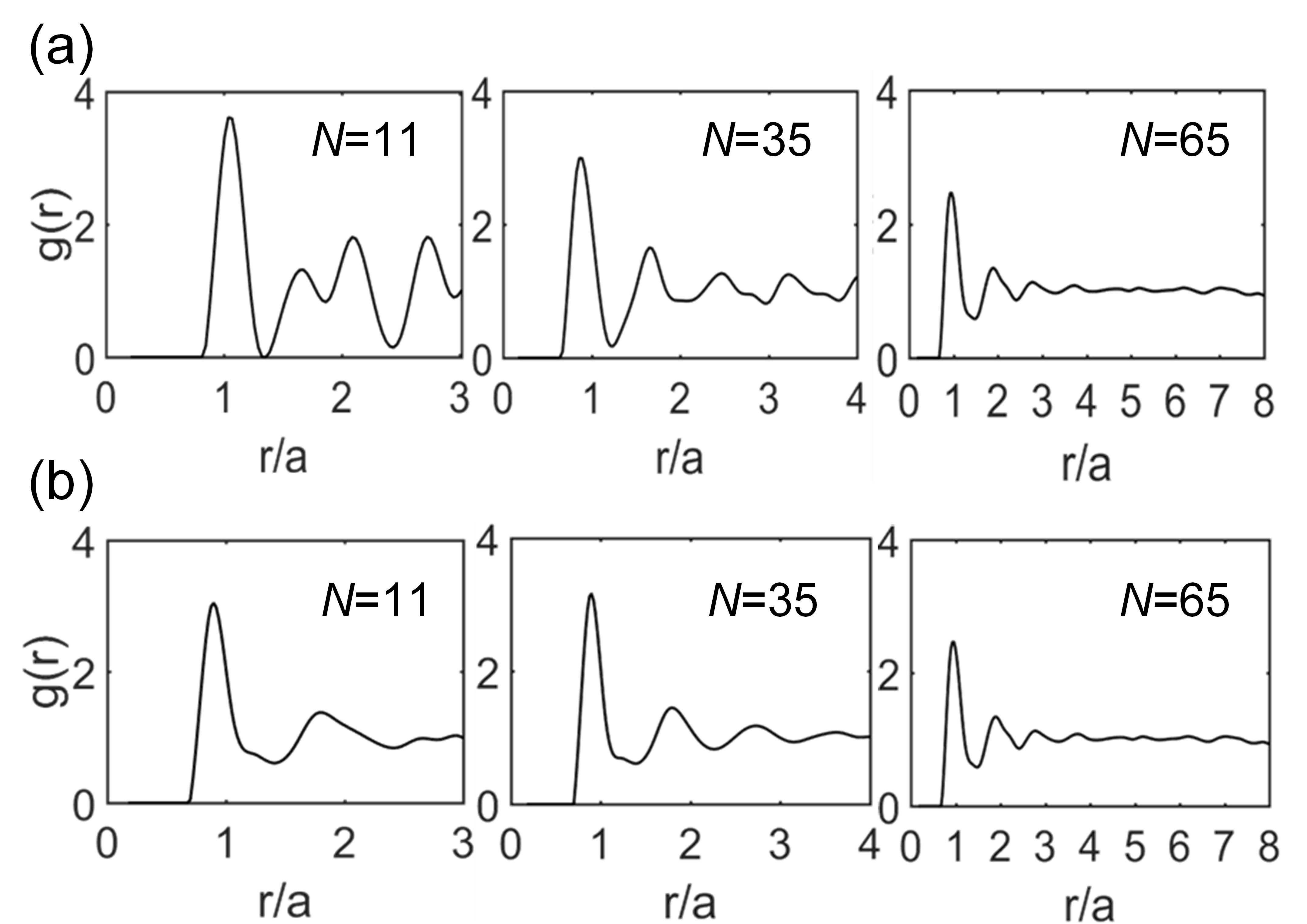}
	\caption{Pair distribution functions $g(r)$ for $\mathbf{(a)}$ SHU structures and $\mathbf{(b)}$ HD structures with different number of nanoholes. {The scaling of the $x$-axis is consistent with the size of the corresponding structures with different numbers of nanoholes. For example, for SHU structures with 11 nanoholes, the size of this configuration is $L$ = $a$$\sqrt{N}$ $\approx$ $3a$. Obviously, there are large fluctuations when $N$ is small for SHU structures while the $g(r)$ can keep relatively steady for HD structures.}  }\label{fig:4} 
\end{figure}

{In fact, size effects have been considered in much of literature especially for solar cells to design a well-designed supercell \cite{Martins2012Engineering}. In the previous work, the characteristics of these ordered or random structures can be easily realized with a small size configuration, for example, a supercell with $3a$$\times$$3a$ was considered in Ref. \cite{Ding2017Design,Peretti2013Absorption}. But for SHU structures with short-range order, there is a strict requirement for the structure factor $S(k<k_{\textup{c}}) = 0$. Larger ${k_{\text{c}}}$ means that a larger supercell should be considered. Therefore, it is significant and necessary to study the size effects firstly. } 
 
Theoretically, the size effects for SHU structures can be analyzed and predicted in real space via the pair distribution function $g(r)$ and simple $k$-space analysis. Figures \ref{fig:4}(a) (SHUs) and \ref{fig:4}(b) (HDs) demonstrates the changes of  $g(r)$  for two structures with representative {small} number{s} of nanoholes. To explain the size effects, the maximum value of horizontal coordinate is given at $x_{\textup{max}}=L/a$, where $a$ is the characteristic length related to number density $\rho$ and yields $a=1/\sqrt{\rho}=\sqrt{f/{(\pi R^{2})}}$. As shown in Fig.~\ref{fig:4}(b), for HD structures, even though $g(r)$ exhibits a small fluctuation within small distance $r$, the value approaches unity at large $r$ because of configurations without long-range order. {On the contrary}, fluctuation peaks appearing in small size of the supercell (Fig. \ref{fig:4}(a)) indicates the development of  pseudo periodic rather than desirable short-range order especially when $N = 11,~35$ for SHU structures. For large cases, the suppression of long-range density fluctuations is apparently visible and the structures have possessed stealthy hyperuniform features with short-range order already.
  
\begin{figure}[t]
	\centering
	\includegraphics[width=0.85\linewidth]{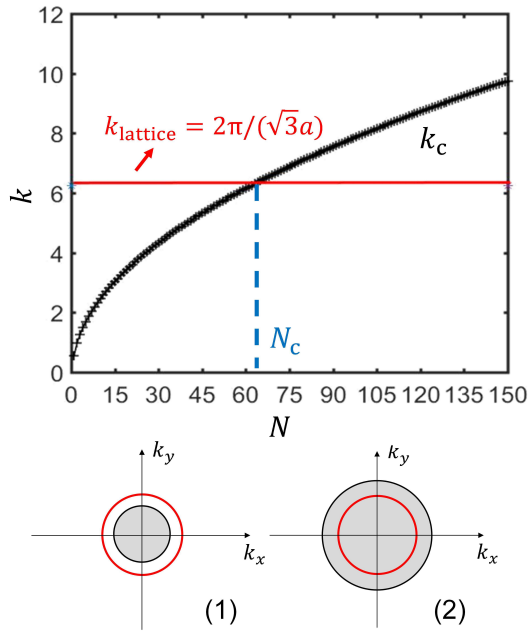}
	\caption{{The black curve is the size of {the} constrained region changing with the number $N$ of holes. The red line is the criterion to find {a} minimum size $N_{\textup{c}}$. In region (1) $N<N_{\textup{c}}$, $k_{\textup{lattice}}>k_{\textup{c}}$, the structures are similar to ordered structures, while in region (2) $N>N_{\textup{c}}$, the structures can be regarded as desirable SHU structures with short-range order because of $k_{\textup{lattice}}<k_{\textup{c}}$. Note that this analysis is merely valid for small cases in our system. For large systems, i.e. $N > 1000$, the size of region $\Omega$ should be well chosen in the scheme.}}\label{fig:5}
\end{figure}

{On the other hand,} these changes in topological structure can be predicted exactly by investigating the relationship between $k_{\textup{c}}$ and the typical wavenumber $k_{\textup{lattice}}$ induced by ordered phase. {We know} that the constrained region $\Omega$ ($k<k_{\textup{c}}$) is influenced by the number of elements $N$ and constrained level $\chi$ {giving} $k_{\textup{c}}=\sqrt{(2M(k_{\textup{c}})+1)/\pi}$. Generally, for a given $\chi=0.5$, we hope to find a minimum value $N_{\textup{c}}$ satisfying that structures with the same constraint parameter possess the same level of short-range order when $N>N_{\textup{c}}$. The main feature of SHU patterns is that some independently wavevectors can be restricted with $S(k)=0$  within $k<k_{\textup{c}}$ in reciprocal space. Thereby, if the wavevectors induced by ordered phase are well constrained in the region $\Omega$, the desirable SHU patterns can be developed. In Fig. \ref{fig:4}(a), there is a pronounced peak satisfying $r=\sqrt{3}a$ in each sub-figure, indicating the development of the ordered phase like hexagonal lattices besides $r=a$. {Then}, we hope that for a desirable SHU pattern, $k_{\textup{lattice}}=2\pi n/(\sqrt{3}a)$ ($n=\pm$1, $\pm$2, $\cdots$) should be well suppressed in region $\Omega$ and meet $k_{\textup{lattice}}<k_{\textup{c}}$. In our systems, $a$ is a constant and $k_{\textup{c}}$ is a function of $N$, therefore, it is easy to find a minimum number $N_{\textup{c}}$ with $2\pi/(\sqrt{3}a)\approx k_{\textup{c}}(N_{\textup{c}})$ when $\rvert n \rvert_{\textup{min}}=1$. As such, for a given $\chi$, when $N<N_{\textup{c}}$, the structures are not real stealthy hyperuniform with obvious long-range order and more like a periodic one. On the contrary, structures with $N>N_{\textup{c}}$ are expected to be stealthy hyperuniform with expected short-range order. 

The corresponding comparison results are illustrated in {Fig.~\ref{fig:5}}.~The red line is marked as the wavenumber $k_{\textup{lattice}}=2\pi/(\sqrt{3}a)$. Above the red line with the minimum number $N_{\textup{c}}=62$ (the blue dot line), desirable SHU structures with a fixed constraint level can be developed. Then, it is understandable that the absorption spectra of SHU structures with $N=11, 35$ are more similar to that of ordered structure compared in Fig. \ref{fig:2}(a). When $N$ is larger than {the} $N_{\textup{c}}$, long range density fluctuation is well suppressed in working waveband ({Fig. \ref{fig:4}(a)}), the absorption spectra of SHU structures become gradually smoother (Fig. \ref{fig:2}(a)) and the integrated absorption keeps steady reasonably ({Fig. \ref{fig:3}(b)}). Therefore, in the next section, a large system should be considered when studying the influence of different levels of short-range order in absorptivity. Inversely, HD patterns are developed in real space and there is no strict restriction with $S(k<k_{\textup{c}})=0$ in reciprocal space.

\section{Influence of the degree of short-range order on absorptivity}

\begin{figure}[t]
	\centering
	\includegraphics[width=0.9\linewidth]{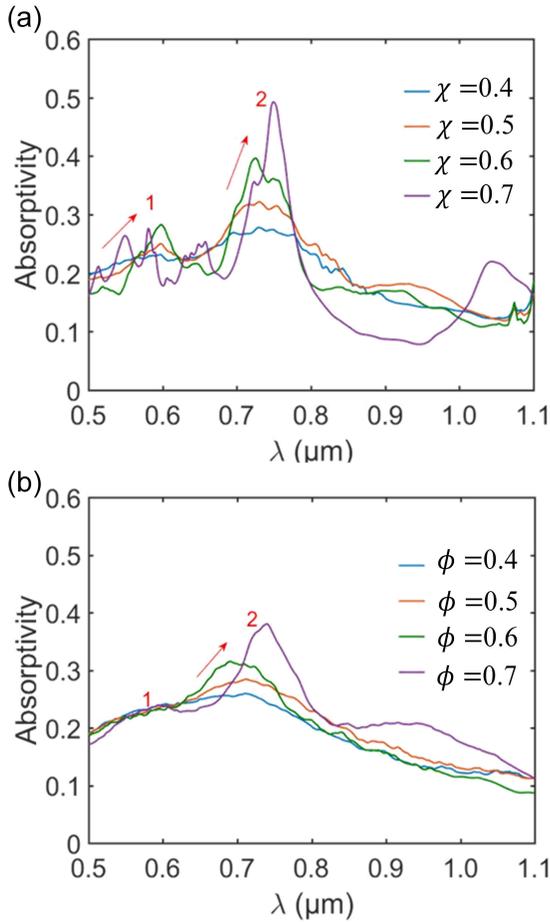}
	\caption{{The absorption spectra of $\mathbf{(a)}$ SHU and $\mathbf{(b)}$ HD structures with different levels of short-range order. {There are two pronounced absorption peaks in both two structures, denoting by No.1 and No.2, respectively.}}}\label{fig:6}
\end{figure}

Even though some studies have investigated the absorptivity in thin-film nanostructures with random distributions, the role of controlled structural correlations in disordered media is seldom discussed. In this section, we consider SHU and HD structures with different levels of short-range order, containing 200 nanoholes for each configuration, which is reliable since the main feature of short-range order can be {obtained}. The comparison results about absorptivity are discussed and the underlying physical mechanisms are also revealed.

Predictably, as shown in Fig.~\ref{fig:6}, gradually sharper absorption peaks are ascribed to an {important} role of local order. Short-range order is able to introduce definite phase correlations of light scattered by adjacent scatterers, which leads to constructive or destructive interference in certain directions and wavelengths. The same trend can be observed in both two structures but the performance of SHU structures shows some pronounced peaks. With the increasing of constraint parameter $\chi$ ranging from 0.4 to 0.7, the interference peak labelled with 1 appears and becomes gradually narrower caused by the minimum distance $a$ between nearest nanoholes. The second peak is the most remarkable and the mechanism behind it can be explained by Bragg's law induced by local ordered phase ({a} hexagonal lattice, $\sqrt{3}a$). As the degree of local order increases, the rising strength of Bragg scattering strengthen the in-plane multiple scattering and enhance the absorption in the slab, which will be {unveiled} in the next section.

\begin{figure}[t]
	\centering
	\includegraphics[width=0.85\linewidth]{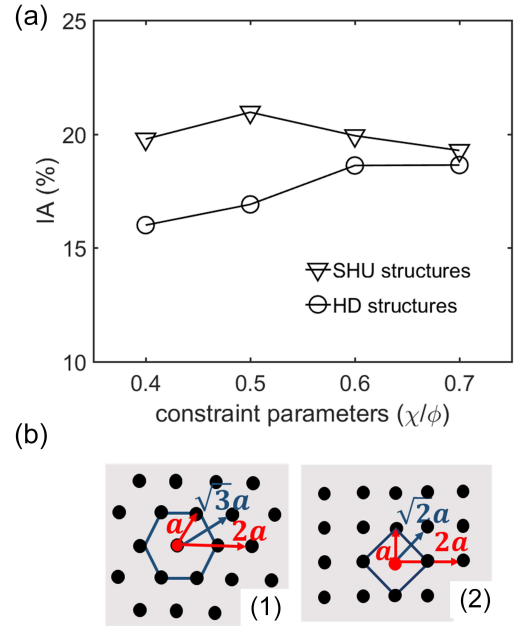}
	\caption{$\mathbf{(a)}$ The integrated absorption of SHU structures and HD structures with different levels of short-range order. $\mathbf{(b)}$ Sketch views of (1) hexagonal lattices and (2) square lattices. The coupled strength decreases during the phase changing from hexagonal lattices to square lattices since the number of nearest scatterer reduces.} \label{fig:7}
\end{figure}

\begin{figure}[htbp]
	\centering
	\includegraphics[width=1\linewidth]{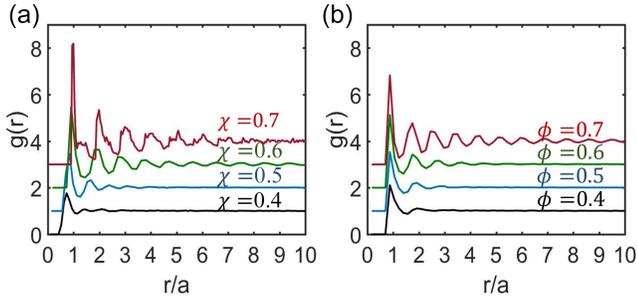}
	\caption{Pair distribution functions $g(r)$ for $\mathbf{(a)}$ SHU structures and $\mathbf{(b)}$ HD structures with different levels of short-range order. With each increase by $\vartriangle$$\chi=\vartriangle$$\phi=0.1$, curves are shifted vertically by 1 for better separation. Each curve is the result of an averaging over 100 statistically independent patterns with the same constraint level.} \label{fig:8}
\end{figure}

Figure \ref{fig:7}(a) illustrates the $IA$ of two disordered structures as a function of the constraint parameters, showing that the integrated absorption of SHU structures are superior to that of HD structures.~Moreover, a turning point appearing at around $\chi=0.5$ indicates that there is an optimal state to achieve a balance between short-rang order and disorder, which enables us to design a variety of media with novel optical properties by merely taking the degree of local positional order into consideration. Actually, this variation also depends on the features of local phase for SHU structures when operating the generation program. SHU structures witness a phase transformation when {the} constraint parameter {is in the range of} $0.5 < \chi < 0.7$ \cite{froufe2016role}, thus {the} local-order phase changes from six-nearest-neighbors to four-nearest-neighbors, resulting in the reduction of the strength of constructive interferences. Such features can be seen clearly from the changes of pair distribution function $g(r)$ illustrated in Fig.~\ref{fig:8}(a). Reasonably, {as shown in both Figs. \ref{fig:8}(a) and (b),} the first peak is induced by the minimum distance between nearest scatterers, and the second peak appearing at distance $r =\sqrt{3}a$, indicates that the local-order phase of structures shows a high similarity with hexagonal lattices (Fig. \ref{fig:7}(b)(1)). When {the} constraint parameter approaches $\chi=0.7$, another obvious peak occurs {for SHU structures} and its position agrees well with {the} lattice constant $(\sqrt{2}a)$ of square lattices (Fig.~\ref{fig:7}(b)(2)), weakening coupling strength between nearest scatterers. Thus, the integrated absorption shows a successive decline with $IA=19.28\%$ slightly smaller than that of when $\chi=0.7$, suggesting that the appearance of square lattices has a negative effect on absorption enhancement for SHU structures.

\begin{figure}[b]
	\centering
	\includegraphics[width=1\linewidth]{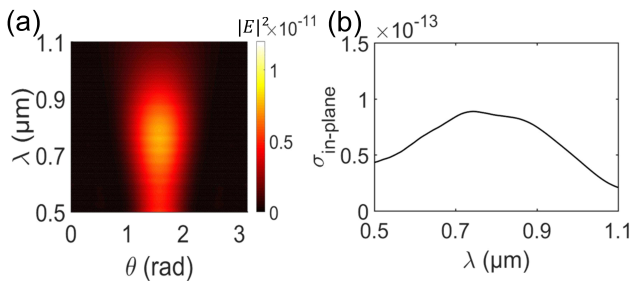}
	\captionsetup{format=default}
	\caption{$\mathbf{(a)}$ The in-plane differential scattering cross section of the individual scatterer corresponding with $\mathbf{(b)}$ the in-plane scattering cross section as a function of wavelength. Note that we just collect the scattering intensity in $x$ and $y$ directions, and there are weak morphology-dependent-like resonances for $\sigma_{\textup{in-plane}}$ appearing around 0.75 $\upmu$m.  } \label{fig:9}
\end{figure}

\section{Scattering theory considering short-range order}

Apparently, the degree of short-range order plays an important role in absorption enhancement in our systems. {Before,} for a disordered system with {a} finite thickness, {Kevin $et$ $al$ have investigated the capability of optical modes excitation by collecting resonances information in a thin film given by spectral function \cite{vynck2012photon,Lagendijk1996Resonant}. This model is powerful when studying the resonance density distributions of random systems illustrating a map of electromagnetic field distributions, but lacks detailed information about the effect of positional correlations of scatterers on dependent scattering. Besides, the effects caused by the individual scatterer or the positional correlations cannot be distinguished clearly as well.} {Therefore, we employ scattering theory} to account for structural correlations and therefore the modifications they induce in the scattering features of the system. 

{Fundamentally, as for a medium where scatterers are randomly arranged and far from each other, the independent scattering approximate is valid. And the total scattering cross section can be simply expressed by $\sigma = \sum_{i=1}^{N} \sigma_{i}$ with $ \sigma_{i}$ being the cross section of the individual scatterer \cite{RevModPhys.23.287}. However, in our systems, the existence of short-range order induced by the typical distance between adjacent scatterers will imply a certain phase relation depending on wavelength and direction of scattered waves, leading to either destructive or constructive interference between them. Therefore, we consider} effective differential scattering cross section $d \sigma^{*}/d \theta$ by correcting the single scatterer differential cross  section $d \sigma/d \theta$ with {the} structure factor $S(q)$ expressed as  \cite{PhysRevB.42.2621,PhysRevLett.65.512}
\begin{equation}
\frac{d \sigma^{*}}{d \theta}=S(q)\frac{d\sigma }{d \theta},
\end{equation}\label{eq:4}
in which $q=(4\pi n_{\textup{e}}/\lambda)\textup{sin}(\theta/2)$, $\theta$ is the scattering angle and $n_{\textup{e}}$ is the effective refractive index according to Maxwell$-$Garnett mixing rule \cite{garnett1906colours,Sihvola1999Electromagnetic}. This method has been extensively investigated in two-dimensional structures and $d \sigma/d \theta$ can be easily calculated from Mie theory of a infinite cylinder with the incident and scattered light in {the} $x$-$y$ plane\cite{Reufer2007Transport,conley2014light,noh2010double}. But, {the thin-film structure considered here is illuminated in the $z$ direction. Therefore, to account for structural correlations in in-plane scattering and absorption enhancement, we merely collect scattering intensity in the $x$-$y$ plane of a single scatterer} (a thin disk embedded with a nanohole).
\begin{figure}[t]
	\centering
	\includegraphics[width=1\linewidth]{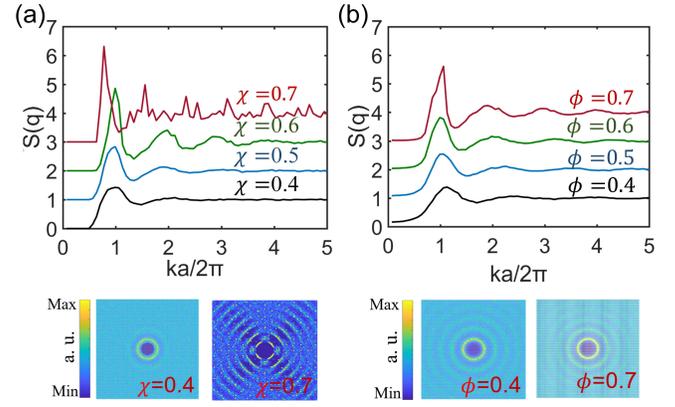}
	\caption{Angular-averaged structure factor of $\mathbf{(a)}$ SHU structures and $\mathbf{(b)}$ HD structures for different levels of short-range order with corresponding 2D structure factor for two typical constraint levels shown on the bottom. With each increase by $\vartriangle$$\chi=\vartriangle$$\phi=0.1$, curves are shifted vertically by 1 for better separation. Each curve is the result of an averaging over 100 statistically independent patterns with the same constraint level. {The in-plane angle ranges 0 $\sim$ 2$\pi$.}} \label{fig:10}
\end{figure}

The angular scattering pattern ($d \sigma/d \theta$) of an individual scatterer is illustrated in Fig.~\ref{fig:9}(a) corresponding with the in-plane scattering cross section with $y$-polarization incidence.~A broad peak appearing in Fig.~\ref{fig:9}(b) {for $\sigma_{\textup{in-plane}}$} indicates the existence of weak morphology-dependent-like resonant modes induced by its geometrical features \cite{Hulst1981Light}. {Since the leakage scattering lights along $z$ direction are useless for in-plane absorption enhancement and structural correlations in the $x$-$y$ plane have no effect on light transport in the third direction, it is reasonable to take in-plane scattering and structural correlations into the aforementioned scattering model} in Eq.~(4). {Then}, we {can} unveil that short-range order will {substantially} modify the resonant scattering patterns in our systems.

Numerically, {for a many-particle system,} the geometrical structure factor {can be accurately calculated via involving collective density variables $C(\mathbf{q})$} {according to Euler's formula} as follows \cite{uche2006collective,torquato2015ensemble,Wu2017Effective}
\begin{equation}
S(\mathbf{q})=\frac{1}{N} \big\|\sum_{j=1}^{N} e^{{-i\mathbf{q}}\cdot{\mathbf{R}_{j}}}\big\|^{2}=1+\frac{2}{N}C(\mathbf{q}),
\end{equation}
in which
\begin{equation}
C(\mathbf{q})=\sum_{j=1}^{N-1} \sum_{i=j+1}^{N} \textup{cos}[\mathbf{q}\cdot(\mathbf{R}_j-\mathbf{R}_i)].
\end{equation}
\begin{figure}[t]
	\centering
	\includegraphics[width=0.9\linewidth]{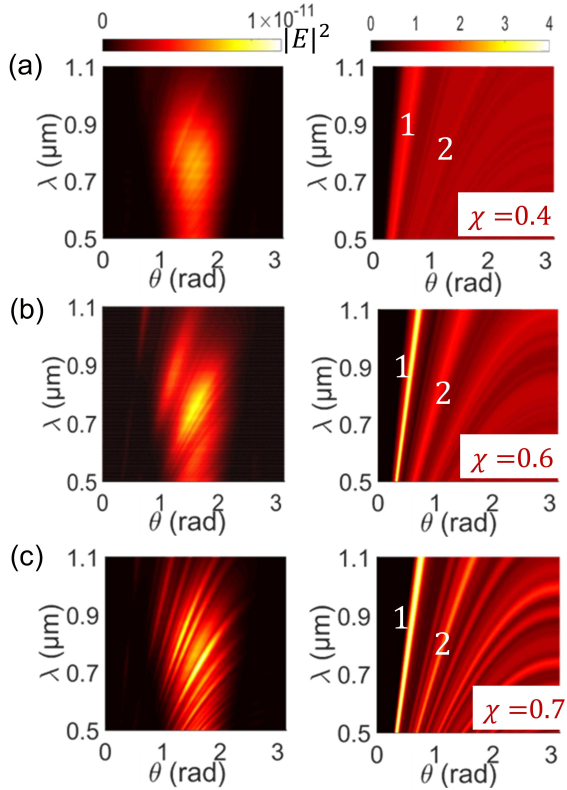}
	\caption{Plots of in-plane effective differential scattering cross section $d \sigma^*/d \theta$ (right) and {the} structure factor $S(\lambda, \theta)$ for SHU structures as a function of $\lambda$ and $\theta$. $\mathbf{(a)}$ $\chi=0.4$, $\mathbf{(b)}$ $\chi=0.6$, $\mathbf{(c)}$ $\chi=0.7$. As the constraint level increases, the second peak induced by typical local-order phase in $S(\lambda, \theta)$ plays an gradually significant role in correcting differential scattering cross section of the individual scatterer. The numbers 1 and 2 indicate the first and second peaks in the structure factor. {We do not put the results of $\chi =0.5$ here because they are similar to results of $\chi=0.4$.} {All patterns share the same color scale bar on the top of each column.}} \label{fig:11}
\end{figure}
The results of angular-averaged structure factor of disordered structures considered in this paper are illustrated in Fig.~\ref{fig:10} with {representative} two-dimensional {contour} plots showing {on the bottom}. {Obviously, the contour plots show circular ring patterns, indicating that the structure is isotropic and there exits dominant wavelength corresponding to radius of the ring \cite{yang2010photonic,liew2011short}. The isotropic scattering also indicates that light scattering in disordered media is insensitive to the polarization states of the incident light \cite{burresi2013two,Ding2017Design}.} {Results show that} positional correlations gradually develop as the $\phi$ and $\chi$ increase with significantly rising height of the peaks in $S(q)$. The second peaks in HD systems {(the second radius in contour plots)} indicate the development of a hexagonal lattice, while new peaks appear for SHU systems showing the formation of a square lattice (see 2D $S(q)$ with $\chi=0.7$), which is consistent with the results of pair distribution function $g(r)$ ({Fig.~\ref{fig:8}(a)}). For large $\phi$ and $\chi$, additional peaks develop meaning the appearance of quasi-long-range order with several pronounced peaks. The positions of these peaks, related to the Bragg's condition \cite{Tin1988Bragg}, can tell us for which scattering vectors we could see an enhancement of scattering. And for {SHU structures} the enhancement will obviously be greater. As such, one can predict that the angular scattering intensity distributions will be influenced significantly due to the development of structural correlations.

\begin{figure}[t]
	\flushleft
	\includegraphics[width=0.9\linewidth]{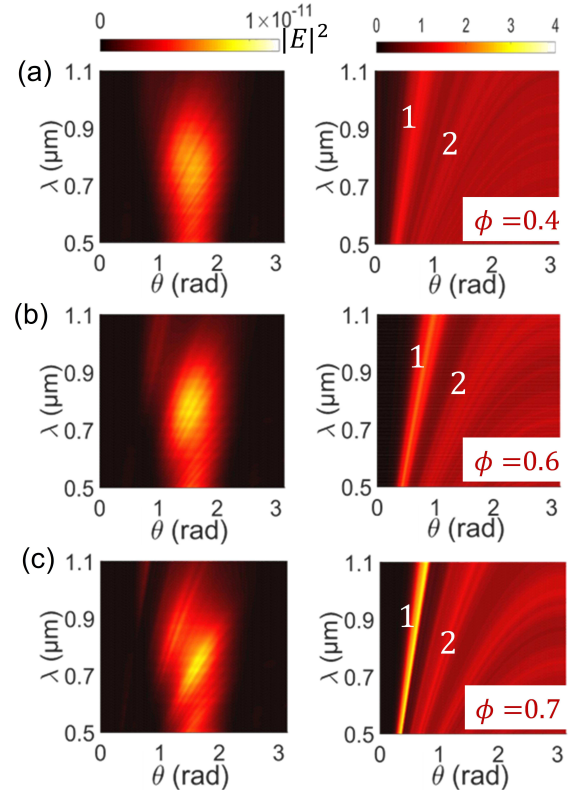}
	\caption{Plots of in-plane effective differential scattering cross section $d \sigma^*/d \theta$ (right) and {the} structure factor $S(\lambda, \theta)$ for HD structures as a function of $\lambda$ and $\theta$. $\mathbf{(a)}$ $\phi=0.4$, $\mathbf{(b)}$ $\phi=0.6$, $\mathbf{(c)}$ $\phi=0.7$. {The numbers 1 and 2 denote the first and second peaks in the structure factor.} {We do not put the results of $\phi =0.5$ here because they are similar to results of $\phi=0.4$.} {All patterns share the same color scale bar on the top of each column.} }\label{fig:12}
\end{figure}

In Fig.~\ref{fig:11} and Fig. \ref{fig:12}, we show the angular and spectral resolved effective differential scattering cross sections $d \sigma^*/d \theta$ {(left)} with the corresponding structure factor $S(\lambda, \theta)$ {(right)} for two disordered structures with different levels of local order. {All patterns share the same color scale bar on the top of each column.} As expected, with higher $\chi$ and $\phi$, $d \sigma^*/d \theta$ exhibits increasingly sharper features in certain wavelengths and angles due to the oscillations of {the} structure factor (see Figs.~\ref{fig:11}(b), (c) and \ref{fig:12}(c)), while there is little change in the cases of small constraint parameters shown in Fig.~\ref{fig:11}(a) and {Figs.~\ref{fig:12}(a), (b).} 

Actually, for the resonances induced by the individual scatterer, the second Bragg-like peak of $S(q)$ (labeled with 2 in Figs. \ref{fig:11} and \ref{fig:12}) plays a dominant role in scattering modification, giving considerable weights to the scattering enhancement. Clearly, for the small constraint condition light scattering between scatterers is independent and the total scattering intensity is a sum of that of each scatterer in a random medium, which is the main reason for the similarity in the shape of {in-plane} scattering cross section (see Fig. \ref{fig:9}(b)) and absorption spectra with small levels of local order (see Fig. \ref{fig:6}). However, for strongly correlated structures, the in-plane scattering is enhanced with gradually pronounced peaks in Figs.~\ref{fig:11}(b), (c) {and 12(c)} resulting from the {interplay} of resonant modes {of the individual scatterer} and Bragg scattering induced by strong local order. Note that extra peaks developing for {SHU} structures with high {degrees of local order} also have a gradual positive impact on scattering for both short and long wavebands, leading to the development of other peaks at either long or short wavelengths in Fig. \ref{fig:6}{(a)}. Besides, it is obvious that the maximum amplitude of scattering intensity and the value of peaks of {the} structure factor for SHU structures are both larger than that of HD structures {with brighter color both in $d \sigma^*/d \theta$ and $S(\lambda, \theta)$}, agreeing well with the results of integrated absorption in Fig. \ref{fig:10}(a). Admittedly, this model also has some limitations, including ignoring the near-field effects of adjacent scatterers, which has a significant impact on effective dielectric index of the surrounding media \cite{liew2011short,Petrova2009,Mcneil2000Multiple}. A more accurate model should be developed by considering near-field effects when study the influence of short-range order in scattering mean free path, scattering anisotropy factor and so on, which will be our further work. 

\section{Conclusion}
To study the impact of short-range order on light absorption in disordered media, we make a detailed comparison between novel stealthy hyperuniform structures with conventional hard disk structures. Firstly, for structures with a given constraint degree, we focus on the absorptivity of supercells and find a {proper} size for SHU structures, thereby possessing desirable short-range order, while there is no strict limitation for HD structures whose structural correlations are insensitive to size effects. Then, for both SHU structures and HD structures, absorptivity can be further tuned by {chang}ing the level of short-range order. {A pronounced peak exits in absorption spectra for both SHU and HD structures {, which can be}} ascribed to {in-plane} Bragg scattering. The analysis of in-plane effective differential scattering cross section correcting by {the }structure factor reveals that higher degree of short-range order leads to stronger Bragg scattering with several peaks in the structure factor, which can well modify the angular and spectral scattering distributions of the individual scatterer and result in better absorption performance in certain waveband. {The results can also be generalized to structures with different thicknesses or based on realistic materials. }

Moreover, although SHU structures and HD structures show high similarities in the development of PBGs in reciprocal space \cite{froufe2016role}, SHU structures are more flexible and show better performance in light absorption manipulation. It indicates that absorption is a macroscopic integrated quantity over the entire real space so that there is no symmetry with respect to real space and reciprocal space. {Notably, light scattering in disordered media is isotropy and independent on the polarization states of incident light, therefore the SHU and HD structures would show definite absorption enhancement for $x$-polarization as well, which has been verified in several research papers \cite{burresi2013two,Ding2017Design}.} The present study opens new opportunities in controlling light absorption using novel disordered nanostructures with short-range order. Furthermore, we expect SHU structures to be desirable in other applications like manipulating the optical performance of random lasing \cite{Cao2013} and structural coloration \cite{Xiaoe1701151}.
\\

\noindent
$\textbf{Funding}.$ The National Natural Science Foundation of China (51636004, 51476097); Shanghai Key Fundamental Research Grant (16JC1403200); The Foundation for Innovative Research Groups of the National Natural Science Foundation of China (51521004).

\bibliography{ref}

\bibliographyfullrefs{ref}

\end{document}